\documentclass[journal=jacsat,manuscript=article]{achemso}

\usepackage{tikz}
\usepackage[version=3]{mhchem}

\usepackage{chemformula} 

\usepackage{graphicx}%
\usepackage{multirow}%
\usepackage{amsmath,amssymb,amsfonts}%
\usepackage{amsthm}%
\usepackage{mathrsfs}%
\usepackage[title]{appendix}%
\usepackage{xcolor}%
\usepackage{textcomp}%
\usepackage{manyfoot}%
\usepackage{booktabs}
\usepackage{array}
\usepackage{caption}
\usepackage{algorithm}%
\usepackage{algorithmicx}%
\usepackage{algpseudocode}%
\usepackage{listings}%
\usepackage{soul}
\usepackage{float}
\usepackage{lineno}
\usepackage{pifont}

\soulregister{\cite}7
\soulregister{\citep}7
\soulregister{\citet}7
\soulregister{\ref}7
\soulregister{\pageref}7

\newcommand{\subsubsubsection}[1]{
    \vspace{0.5em}
    \noindent\textbf{#1}
    \vspace{0.2em}
}

\title{Chemistry-Enhanced Diffusion-Based Framework for Small-to-Large Molecular Conformation Generation}

\author{Yifei Zhu}
\affiliation{SCNU Environmental Research Institute, Guangdong Provincial Key Laboratory of Chemical Pollution and Environmental Safety, School of Environment, South China Normal University, Guangzhou 510006, P. R. China.}
\alsoaffiliation{MOE Key Laboratory of Environmental Theoretical Chemistry, South China Normal University, Guangzhou 510006, P. R. China.}
\author{Jiahui Zhang}
\affiliation{SCNU Environmental Research Institute, Guangdong Provincial Key Laboratory of Chemical Pollution and Environmental Safety, School of Environment, South China Normal University, Guangzhou 510006, P. R. China.}
\alsoaffiliation{MOE Key Laboratory of Environmental Theoretical Chemistry, South China Normal University, Guangzhou 510006, P. R. China.}
\author{Jiawei Peng}
\affiliation{MOE Key Laboratory of Environmental Theoretical Chemistry, South China Normal University, Guangzhou 510006, P. R. China.}
\author{Mengge Li}
\affiliation{MOE Key Laboratory of Environmental Theoretical Chemistry, South China Normal University, Guangzhou 510006, P. R. China.}
\author{Chao Xu}
\affiliation{SCNU Environmental Research Institute, Guangdong Provincial Key Laboratory of Chemical Pollution and Environmental Safety, School of Environment, South China Normal University, Guangzhou 510006, P. R. China.}
\alsoaffiliation{MOE Key Laboratory of Environmental Theoretical Chemistry, South China Normal University, Guangzhou 510006, P. R. China.}
\author{Zhenggang Lan}
\affiliation{SCNU Environmental Research Institute, Guangdong Provincial Key Laboratory of Chemical Pollution and Environmental Safety, School of Environment, South China Normal University, Guangzhou 510006, P. R. China.}
\alsoaffiliation{MOE Key Laboratory of Environmental Theoretical Chemistry, South China Normal University, Guangzhou 510006, P. R. China.}
\email{zhenggang.lan@m.scnu.edu.cn}

\begin{document}
\graphicspath{{Figures/}{Figures/failed_molecules/}}

\abstract{
    Obtaining 3D conformations of realistic polyatomic molecules at the quantum chemistry level remains challenging, and although recent machine learning advances offer promise, predicting large-molecule structures still requires substantial computational effort.
    Here, we introduce StoL, a diffusion model-based framework that enables rapid and knowledge-free generation of large molecular structures from small-molecule data.
    Remarkably, StoL assembles molecules in a LEGO-style fashion from scratch, without seeing the target molecules or any structures of comparable size during training.
    Given a SMILES input, it decomposes the molecule into chemically valid fragments, generates their 3D structures with a diffusion model trained on small molecules, and assembles them into diverse conformations.
    This fragment-based strategy eliminates the need for large-molecule training data while maintaining high scalability and transferability.
    By embedding chemical principles into key steps, StoL ensures faster convergence, chemically rational structures, and broad configurational coverage, as confirmed against DFT calculations.
}

\maketitle

\section{Introduction}
    Molecular conformation, defined as the precise three-dimensional (3D) arrangement of atoms within a molecule, profoundly influences its chemical and physical properties, including reactivity, stability, and intermolecular interactions. \cite{seeman1983effect,hawkins2017conformation,fang2014conformational,linclau2022relating}
    As the molecular size grows, the number of accessible conformations grows exponentially, making exhaustive exploration of the conformational space computationally challenging.
    However, understanding molecular conformational distribution is essential, because it provides key insights into molecular behaviors and functions across a variety of applications, such as biological sciences, \cite{tang2024recent} drug design \cite{lyne2002structure,anderson2003process,yu2016computer,mcnutt2023conformer}, and catalyst development \cite{freeze2019search}, and so on.

    Stable molecular configurations are typically identified through geometry optimization, provided reasonable initial guess structures are available.
    Traditional methods for generating these initial geometries fall into two primary categories.
    Physics-based approaches, such as metadynamics and molecular dynamics, offer high accuracy by explicitly exploring the configurational space but incur significant computational costs, limiting their ability to comprehensively identify low-energy conformations. \cite{laio2002escaping, pracht2020automated,labute2010lowmodemd,bedoyaMoltiverseMolecularConformer2025}
    On the other hand, heuristic-driven methods leverage empirical rules and fragment libraries to rapidly generate plausible conformers, while computationally efficient, these approaches often sacrifice precision, frequently overlooking critical states. \cite{rdkit, riniker2015better,hawkins2010conformer,seidelHighQualityConformerGeneration2023,Frog22010}
    As a summary, this inherent trade-off between computational efficiency and physical accuracy remains a central challenge in computational chemistry.


    In recent years, machine learning (ML) methods begin to bridge this gap, yielding preliminary advancements. \cite{Simm2020,pmlr-v139-xu21f,ganeaGeoMolTorsionalGeometric2021,zhu2022direct,chengScalableAutoregressive3D2025}
    Among these, diffusion models, initially developed for high-fidelity generative tasks in image and video synthesis \cite{hoDenoisingDiffusionProbabilistic2020,rombachHighResolutionImageSynthesis2022a,hoVideoDiffusionModels2022,dalalOneMinuteVideoGeneration2025,dhariwalDiffusionModelsBeat2021,diffusionreview2023} attracted increasing attention for their ability to model complex data distributions.
    As one of the most powerful generative paradigm, diffusion models show particular promise for molecular structure generation.
    Preliminary applications in chemistry demonstrated their potential, \cite{wangDiffusionModelsMolecules2025, pmlr-v162-hoogeboom22a,zhangSDEGenLearningEvolve2023, kimDiffusionbasedGenerativeAI2024,fengUniGEMUnifiedApproach2025,duanOptimalTransportGenerating2025,morehead2024geometry,schneuingStructurebasedDrugDesign2024, yimDiffusionModelsProtein2024,niPretrainingFractionalDenoising2024,igashovEquivariant3DconditionalDiffusion2024,huangDualDiffusionModel2024}
    including early efforts in molecular conformation generation. \cite{xuGeoDiffGeometricDiffusion2022,torsionaldiffusion_jing,liuNEXTMOL3DDIFFUSION2025,leeDiSCODiffusionSchrodinger2024,morehead2024geometry}

    Although significant advances were made, current available diffusion-based approaches for conformation generation of polyatomic molecules in realistic chemical settings face several limitations.
    For instance, many existing approaches depend on rather complex architectures that are computationally intensive to train and infer, particularly as molecular size and dimensionality grow, undermining scalability.
    Due to purely data-driven nature, most approaches lack embedded chemical principles, and thus result in substantial computational cost and slower training processes.
    Consequently, these models demand large, high-quality datasets for training, which are often infeasible for polyatomic compounds due to the challenges of exhaustive conformer sampling.
    Moreover, the absence of embedded physical and chemical constraints frequently results in the generation of unphysical or energetically implausible structures, even among otherwise high-performing models.
    A growing body of evidence demonstrates that incorporating domain-specific knowledge can dramatically improve both efficiency and reliability. \cite{Keith2021,karniadakisPhysicsinformedMachineLearning2021,zhangEmbeddedAtomNeural2019, pinn2019, stuyverQuantumChemistryaugmentedNeural2022,ullahPhysicsinformedNeuralNetworks2024a,yangCrossModalPredictionSpectral2024,cignoni2024electronic}
    Additionally, many workflows depend on external software for tasks such as hydrogen addition or generating starting geometries
    \cite{torsionaldiffusion_jing,liuNEXTMOL3DDIFFUSION2025,leeDiSCODiffusionSchrodinger2024}
    introducing overhead and hindering seamless end-to-end generative pipelines.
    Overcoming these limitations is essential to establishing diffusion models as practical, deployable tools for real-world chemical applications.

    To address these challenges, we present StoL (Small-to-Large), a novel end-to-end framework for generating diverse, high-quality conformations of large molecules.
    The StoL framework integrates several key components, each contributing distinct advantages.
    It accepts SMILES strings as input and directly outputs multiple conformation's Cartesian coordinates, streamlining operation and enhancing accessibility.
    Inspired by a LEGO-like modular design, StoL operates through three core steps: fragmentation, generation, and assembly.
    Each step mirrors the process of building with LEGO bricks, ensuring systematic and scalable conformation construction.
    Fragmentation decomposes the input SMILES into smaller, manageable fragments using efficient heuristic rules, akin to selecting LEGO bricks.
    Generation employs a chemically enhanced diffusion model to produce plausible 3D configurations of each fragment, ensuring the LEGO bricks are structurally diverse and chemically valid.
    Finally, the assembly step integrates these fragments into the complete 3D molecular structure, applying chemistry-constrained filtering to eliminate unphysical geometries, analogous to assembling LEGO pieces into a stable, accurate model.
    This integrated workflow offers several important benefits.
    The black-box, end-to-end design requires no expert intervention, making the framework highly easy to use.
    More importantly, the modular LEGO-style architecture fundamentally circumvents the need for large-molecule databases—a major limitation of traditional methods by leveraging a ``small-to-large'' strategy.
    By embedding chemically informed rules and constraints at multiple stages, StoL transcends purely data-driven approaches, ensuring not only enhanced predictive accuracy but also chemically plausible and thermodynamically meaningful conformations.

    A central feature of the StoL is the explicit incorporation of chemical principles into the current diffusion-based framework,
    which significantly improve the efficiency and performance of both training and prediction.
    Chemical knowledge is explicitly integrated into key stages of the StoL protocol.
    First, a two-step training strategy is designed for the diffusion model, consisting of a purely data-driven stage followed by a chemistry-enhanced (CE) stage.
    The CE stage incorporates the Sinkhorn \cite{sinkhorn1967concerning} and Gumbel-softmax \cite{jangCategoricalReparameterizationGumbelSoftmax2017} algorithms, along with planarity checks for atomic ring systems.
    These additions guide the model to learn chemically meaningful patterns, remarkably improving training efficiency and overall performance.
    Second, during the assembly phase, several chemoinformatics-inspired constraints are applied to remove nonphysical 3D structures, essentially ensuring the chemical validity of all generated conformations.
    As the consequence, StoL's LEGO-inspired architecture facilitates the generation of isomers across a broader conformational space, akin to the extensive exploration achieved through molecular dynamics, as validated by \textit{ab initio} calculations.
    Overall, our results demonstrate that StoL efficiently generates high-quality conformations for polyatomic molecules in a fully end-to-end manner.
    This work highlights the critical role of integrating fundamental chemical principles into ML models for effective chemical space exploration.
    The success of our approach demonstrates that embedding domain-specific chemical knowledge is crucial for the practical application of these models in real-world chemical systems, providing valuable insights for future advancements in physics-informed molecular generation.


\section{Results}

\subsection{A Brief Description of StoL}

\begin{figure}[H]
    \centering
    \includegraphics[width=0.7\linewidth]{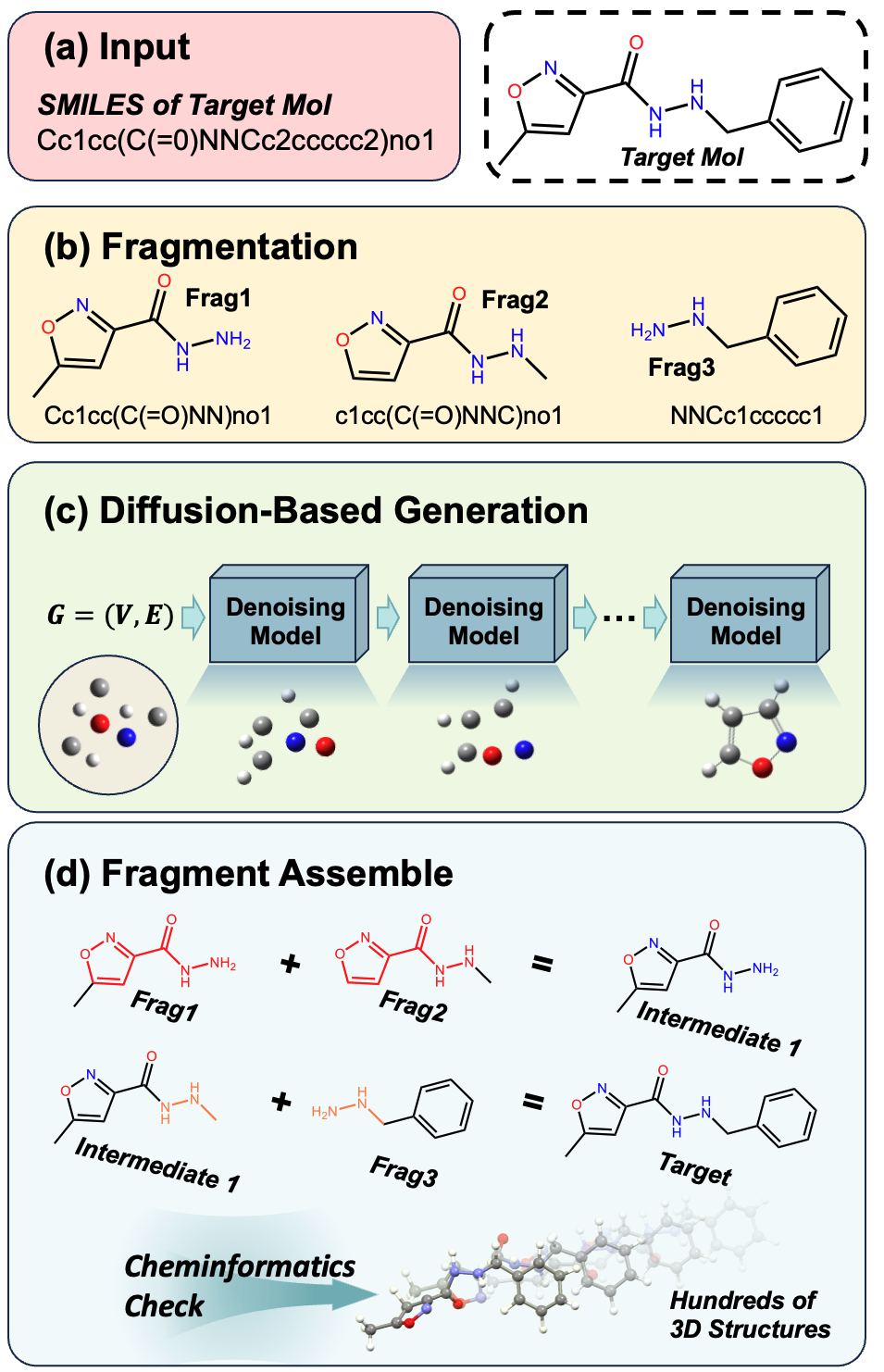}
    \caption{
    (a) Input SMILES of the target molecule.
    (b) Fragmentation into smaller components.
    (c) Diffusion-based generation using denoising models.
    (d) Fragment assembly to reconstruct the target molecule, followed by chemoinformatics check.
    All molecular geometries were visualized via PyMOL. \cite{pymol}
    }
    \label{fig:fig1}
\end{figure}


    In this section, we introduce an overview of StoL, a chemistry-enhanced diffusion-based ML framework designed to generate diverse, high-quality 3D conformations for large molecules.
    The overall workflow, illustrated in Figure \ref{fig:fig1},
    takes a SMILES string as input and produces an ensemble of chemically and geometrically plausible 3D structures.
    The StoL framework operates through three sequential core stages:
    \begin{itemize}
    \item \textbf{Input processing and molecular fragmentation:} The process begins with a SMILES string, which is systematically decomposed into smaller, chemically valid fragments using predefined heuristic rules.
    This step yields a set of SMILES strings representing the fragments, serving as the foundational building blocks for subsequent conformation generation.

    \item \textbf{Fragment conformation generation:}
    Each fragment's SMILES is converted into a molecular graph.
    A chemistry-enhanced diffusion model then generates multiple plausible 3D conformations for each fragment.
    These conformations undergo rapid geometric filtering, guided by chemical principles, to eliminate unrealistic structures and minimize redundancy.

    \item \textbf{Global assembly and conformation output:}
    The filtered 3D fragments are assembled into complete molecular structures.
    A final chemistry-constrained validation step ensures the structural integrity and thermodynamic stability of each assembled conformation.
    The framework outputs a diverse set of distinct 3D conformers as its final result.

    \end{itemize}
    This modular, end-to-end workflow effectively tackles the challenge of scalable conformation generation for large, flexible molecules, combining computational efficiency with chemical fidelity.
    Following the overview of the StoL framework, we provide a detailed description of its core stages below.

\subsubsection{Input Processing and Molecular Fragmentation}

    The first stage of the StoL framework systematically decomposes each target molecule into several smaller yet chemically meaningful fragments.
    In essence, this step disassembles a complex LEGO structure into its fundamental bricks, facilitating subsequent reconstruction.
    Both inputs and outputs at this stage are represented as SMILES strings, chosen for their broad applicability and ease of handling.

    Given an input SMILES, the algorithm first identifies all single bonds in the molecular graph, excluding those located within ring systems or bonded to hydrogen atoms.
    These eligible bonds are then randomly cleaved to generate molecular fragments.
    In practice, only those fragments containing 6-10 heavy atoms (C, N, O, F) are retained, with this upper bound adapted to the molecular size distribution in the training dataset.
    The total number of fragments is also limited to prevent over-fragmentation; if no valid combination satisfies these constraints, they are gradually relaxed to ensure successful reconstruction.
    The resulting fragments are sorted in descending order of size to prioritize larger substructures.

    Next, a largest-fragment-first traversal strategy is employed to select a set of chemically valid fragments from the pre-generated pool that already satisfies the size constraint (6-10 heavy atoms).
    The selection follows three heuristic rules to ensure chemical validity and completeness:
    (\romannumeral1) adjacent fragments must share at least three heavy atoms to maintain structural coherence;
    (\romannumeral2) all selected fragments, when combined, must fully reconstruct the original molecule; and
    (\romannumeral3) the number of fragments is balanced to optimize computational efficiency and minimize error accumulation during subsequent assembly.

    The implementation resembles a tree traversal process.
    First, the algorithm designates the first atom in the SMILES string as Atom-Zero.
    The largest fragment containing Atom-Zero is selected as the root node, denoted as \textit{Frag1}.
    Subsequent fragments are iteratively attached under the predefined constraints. When multiple candidates satisfy the selection criteria, the largest fragment is preferred; if several have identical sizes, one is randomly chosen.
    Following this rule, the algorithm identifies the next fragment (\textit{Frag2}) by examining all fragments that share a common region with \textit{Frag1}.
    The procedure repeats until a valid fragment set is obtained, efficiently exploring the combinatorial space while avoiding redundancy.

    This stage ultimately yields a collection of SMILES strings representing chemically valid fragments, which serve as the foundational building blocks for the subsequent 3D conformation generation step.

    \subsubsection{Fragment Conformation Generation}

    In the second stage of the StoL framework, a ML model generates 3D structures for molecular fragments, serving as the core of the conformation generation process. Here the diffusion-based generative model is used, in which the fragment-level SMILES strings are taken as input and a few of 3D structures for each fragment are generated as the output. In another words, all essential LEGO building blocks are generated.

\begin{figure}[H]
    \centering
    \includegraphics[width=0.8\linewidth]{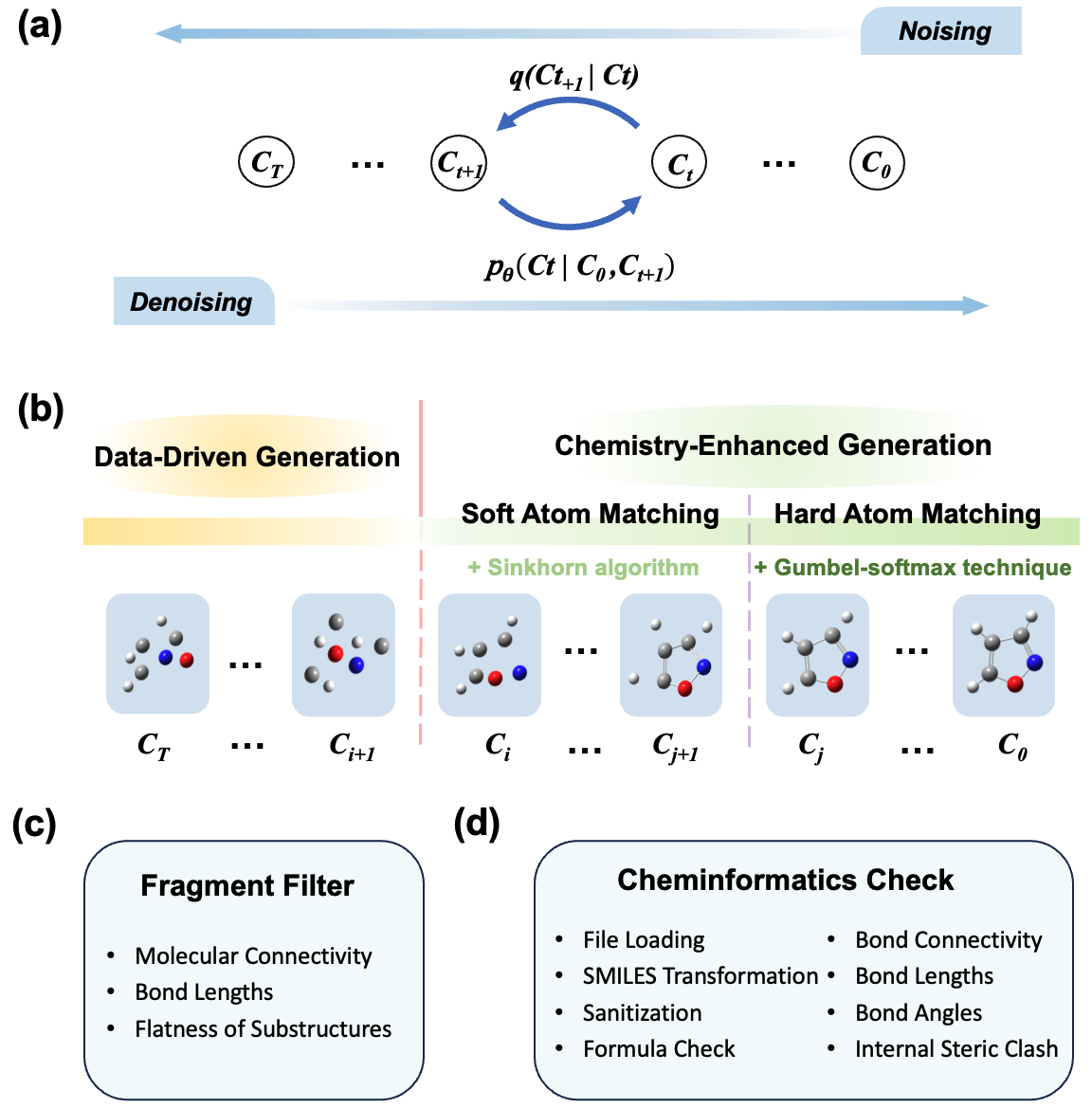}
    \caption{
    (a) Schematic illustration of the standard diffusion model.
    (b) Diagram of the chemistry-enhanced training strategy.
    (c) Basic validation procedures for fragment generation.
    (d) Chemoinformatics-based validation of assembled molecular structures.
    }
    \label{fig:fig2}
\end{figure}

\subsubsubsection{\textbullet~Model architecture}
\newline
    In principle, our model is built upon the stochastic denoising diffusion framework \cite{hoDenoisingDiffusionProbabilistic2020}.
    Within this framework, model learns a continuous matching score that guides noisy atomic coordinates toward their ground-truth geometries through a gradual denoising process.
    Here, StoL takes fragment-level SMILES strings as input, which are transformed into molecular graphs $G=(V,E)$ encoding atomic and bonding information.
    The diffusion model thus learns to estimate the denoising direction conditioned on $G$, enabling the recovery of accurate 3D structures from randomly perturbed coordinates, as illustrated in Figure~\ref{fig:fig2}.(a).

    To implement this conditional denoising process, StoL employs a SchNet-based graph neural network (GNN) as the denoising backbone \cite{kimDiffusionbasedGenerativeAI2024, schnet2018}, which predicts the noise given the molecular graph $G$ at each diffusion step.
    SchNet offers a computationally efficient and physically grounded architecture that models continuous interatomic interactions via learned filters.
    Conditioned on the molecular graph, it predicts the denoising direction at each diffusion step, iteratively refining atomic coordinates while preserving chemically meaningful relationships.
    This design makes StoL particularly well-suited for fragment-level structure generation in large and complex molecular systems.

\subsubsubsection{\textbullet~Training strategy}
\newline
    In practice, the diffusion model was trained based on a subset of the QCDGE dataset \cite{zhuQuantumChemistryDataset2024}, which contains 438,142 small molecules (6-10 heavy atoms of C, N, O, and F) optimized at the B3LYP/6-31G*/BJD3 DFT level.
    In this dataset, the SMILES strings and the corresponding 3D structures were taken for the training purpose.

    In typical generative models in chemistry, atoms are often treated as rigid spheres diffusing toward target coordinates. \cite{wangDiffusionModelsMolecules2025}
    However, in molecular systems, the relative positions of atoms are physically more meaningful than absolute coordinates.
    For instance, any two atoms with the same element in identical chemical environments are often be considered indistinguishable.
    Therefore, purely data-driven ML model normally overemphasize chemically irrelevant patterns, significant reducing efficiency and accuracy.
    To improve the performance and reduce the computational cost, StoL introduces the two-phase training strategy, as shown in Figure~\ref{fig:fig2}.(b).

    The first phase is still in the data-driven manner, focusing on learning general spatial correlations.
    In this phase, the molecular scaffold is formed from a disordered initial state via standard diffusion training.
    Once a rough geometry was formed, the second phase is invoked, which is chemistry-enhanced by emphasizing chemically meaningful geometric relationships.
    In this phase, we explicitly account for molecular symmetries by considering translation, rotation, mirror, and atom permutation.
    Translational and rotational variance is removed using the Kabsch algorithm \cite{kabsch1976solution, kabsch1978discussion}, while mirror symmetry is resolved by flipping the signs of Z-coordinates and selecting the configuration with the lower RMSD.
    Because the Kabsch alignment assumes consistent atomic ordering, permutation symmetry must also be addressed.
    To achieve this, a soft-matching stage employing the Sinkhorn algorithm, \cite{sinkhorn1967concerning} with the process applied individually to the atoms of each element.
    At the final stage of the CE phase, we introduced a Gumbel-softmax technique \cite{jangCategoricalReparameterizationGumbelSoftmax2017} to enforce a one-to-one correspondence in atom ordering by selecting the maximum entries, ensuring fixed atom order and smooth gradient flow during training.
    In addition to symmetry alignment, two further refinements are incorporated to improve the chemical fidelity of the generated structures.
    Two additional tricks were introduced to improve the chemical accuracy of the generative models.
    (\romannumeral1) A planarity-aware term is introduced into the loss function to prevent aromatic rings from deviating from planarity.
    (\romannumeral2) The weights of hydrogen atoms follow a logarithmic schedule, allowing the model to first focus on the heavy-atom framework.

    All above designs enable StoL to progressively refine molecular structures from coarse to precise, prioritizing chemically relevant atomic relationships and thereby ensuring both geometric fidelity and chemical validity.
    This dual-phase training strategy ensures the model progressively refines fragment conformations from coarse to precise in the diffusion process, focusing on chemically meaningful spatial relationships.
    Compared with purely data-driven approaches, it significantly improves geometric fidelity and chemical validity.

\subsubsubsection{\textbullet~Inference and fragment screening}
\newline
    After the construction of the diffusion model, StoL generates realistic 3D fragment conformations of each chosen fragment through the iterative denoising processes.
    Similar to the training process, the inputs are molecular graph generated from the corresponding SMILES string for the target compound.

    Starting from random noise, dozens of 3D geometries (e.g., 50) for a fragment are generated and refined by a learned score function that guides the iterative updating of atomic positions toward chemically plausible structures.
    These structures are typically generated through several thousand denoising steps (e.g., 5000).
    Thanks to the efficiency of the SchNet backbone and fragment-level learning, the overall computational cost still remains very low compared with exhaustive DFT-based conformational searches required for large molecules.

    After generation, the resulting geometries undergo basic quality assessments, including checks for atomic connectivity and planarity consistency, as shown in Figure~\ref{fig:fig2}.(c).
    Fragments that pass these criteria are retained as qualified ``LEGO bricks'' for the subsequent molecular assembly stage.

    Additionally, to further reduce redundant calculations, dimensionality reduction and clustering techniques are employed, selecting several representative conformations per fragment.
    To reduce the computational cost, we normally choose 10 conformations here, which in principle are enough to cover the rather board distribution of the 3D structures for each fragment with 6-10 heavy atoms.
    This streamlined approach yields a diverse set of high-quality, chemically plausible conformations for the entire molecule.

\subsubsection{Global Assembly and Conformation Output}
    Following the generation of chemically valid fragment conformations, the third stage of the StoL framework focuses on assembling these LEGO-like building blocks into complete molecular structures.
    In this step, adjacent fragments are systematically merged sequentially, starting from the first fragment and progressively incorporating subsequent ones by aligning their shared regions.
    At the end, several chemically plausible 3D structures are finally obtained.

    Specifically, Starting from the \textit{Frag1} and \textit{Frag2} defined in the above section, the process begins by combining them: all possible conformations of \textit{Frag1} are paired with those of \textit{Frag2}, and if the root-mean-square deviation (RMSD) of all common atoms between a given pair falls below a predefined threshold (0.2 \AA \ in this work), the fragments are merged into a combined structure (\textit{Frag1+2}).
    The resulting ensemble of merged \textit{Frag1+2} structures is then used as the base to incorporate the third fragment.
    This recursive procedure continues until all fragments are integrated, ultimately constructing the full set of possible molecular conformations.

    To ensure chemical validity and eliminate unphysical structures that may result from merging only adjacent fragments, the chemistry-enhanced protocol is applied during full-molecule assembly.
    Several cheminformatics checks (Figure~\ref{fig:fig2}.(d)), inspired by PoseBusters \cite{buttenschoen2024posebusters}, are employed to verify the physical plausibility of the assembled structures.
    These checks consist of:
    (\romannumeral1) Basic validation: structural file loading, structure-to-SMILES conversion, SMILES sanitization, and molecular formula verification;
    (\romannumeral2) Geometric validation: bond connectivity, bond lengths and angles, and detection of internal steric clashes across the full conformation.
    If an insufficient number of valid geometries are obtained, additional fragment conformations are retrieved and assembled until adequate coverage is achieved.

In summary, the modular StoL framework show several significant advantages.
First, its LEGO-inspired three-stage workflow streamlines conventional generative modeling by requiring coverage only of the chemical space of small molecular fragments for both training and inference. In contrast, traditional approaches typically rely on the curation of large, diverse molecular datasets. StoL circumvents this limitation, efficiently bridging fragment generation and full-molecule assembly without sacrificing structural diversity.
Second, the integration of chemical principles throughout training, generation, and assembly ensures both computational efficiency and chemical fidelity.
This results in more physically realistic 3D structures and significantly enhances the performance of ML models when addressing complex, real-world molecular problems.
Third, assembling molecules from fragment-level minimum structures enables a systematic exploration of the local potential energy landscape, allowing the generation of a diverse ensemble of accessible conformations.
Finally, by relying solely on SMILES as input, StoL provides an end-to-end, black-box workflow that is highly user-friendly.
Together, these features make StoL particularly suitable for large-scale molecular modeling where data scarcity and geometric complexity remain major challenges.
Importantly, the StoL framework offers considerable flexibility, allowing its configurations to be tailored to the available computational resources, research goals, and dataset properties.
Remaining limitations and potential improvements are discussed in subsequent sections, and further methodological details are provided in the Supplementary Information.




\subsection{Model Configuration and Evaluation Metrics}

    To systematically investigate the impact of different training strategies, we trained three models:
    \begin{itemize}
        \item P-CE-StoL (chemistry-enhanced training with planarity-aware loss);
        \item CE-StoL (chemistry-enhanced training without planarity loss);
        \item non-CE-StoL (without chemistry-enhanced training).
    \end{itemize}
    Additional details of these models provided in the Supporting Information.
    All models were trained under identical settings, including comparable numbers of training epochs.
    Among them, P-CE-StoL, integrating both the chemistry-enhanced (CE) training scheme and planarity-aware loss, is supposed to be the most excellent model that is the reference for the subsequent analyses.

    To evaluate the quality of the generated conformational space, we employed a Boltzmann-weighted RMSD (BRMSD), which provides a more chemically meaningful measure than the conventional RMSD.
    For each reference conformation, the minimum RMSD among all generated structures was calculated and weighted by its Boltzmann factor, derived from the relative energy of the reference conformation.
    The BRMSD therefore represents the energy-weighted average of these minimum RMSD values, emphasizing agreement with low-energy, chemically relevant conformations.
    This metric enables intuitive assessment of the overall geometric quality of the generated ensemble while incorporating thermodynamic relevance.
    A detailed mathematical definition and computational procedure of BRMSD are provided in the Supporting Information.

\subsection{Illustrative Example of StoL Workflow}

\begin{figure}[H]
    \centering
    \includegraphics[width=0.8\linewidth]{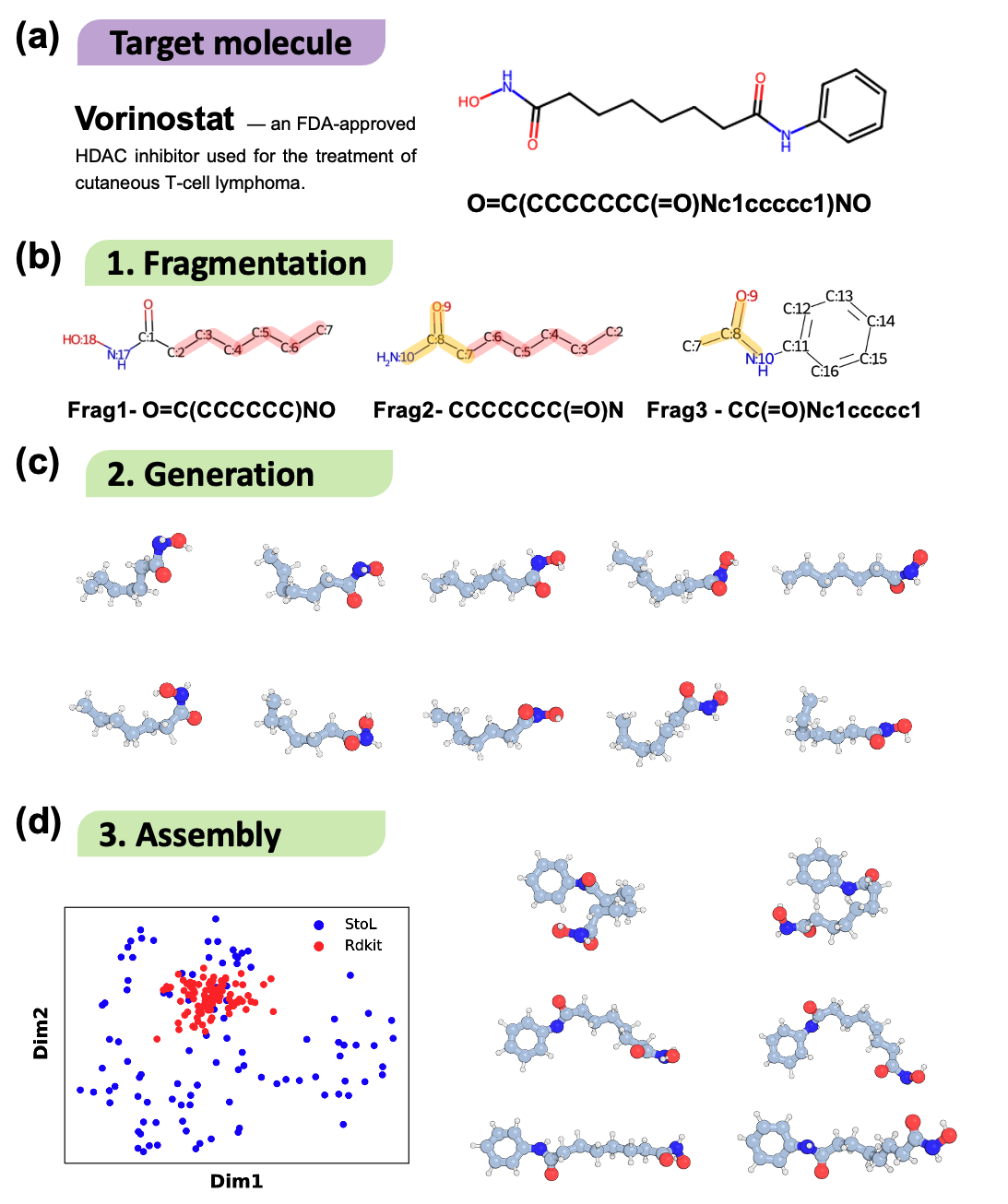}
    \caption{
        Illustrative example of the StoL process applied to vorinostat.
        (a) SMILES representation and molecular structure of vorinostat.
        (b) Fragmentation of the molecule with corresponding fragment SMILES.
        (c) For fragment 1, dimensionality reduction followed by clustering techniques yields 10 representative conformations.
        (d) Final assembly of vorinostat, where six representative 3D structures passed cheminformatics validation check are shown.
    All molecular geometries were visualized via PyMOL. \cite{pymol}
    }
    \label{fig:example}
\end{figure}

    In this section, we demonstrate the effectiveness of the StoL process by a typical example.

    For illustration, we randomly selected the SMILES string \url{O=C(CCCCCCC(=O)Nc1ccccc1)NO}, which corresponds to vorinostat (\ch{C14H20N2O3}, Molecular Weight 264.32), a famous histone deacetylase (HDAC) inhibitor. \cite{yudaev2025recent}
    As shown in Figure~\ref{fig:example}.(a), vorinostat contains multiple pharmacologically relevant functional groups, including hydroxamic acid, amide, and aromatic moieties, underscoring its chemical and biomedical significance.
    Meanwhile, its long and highly flexible aliphatic chain introduces numerous rotatable bonds and intramolecular interactions, making it particularly challenging for conventional conformational exploration methods, which often struggle to efficiently sample such a broad and flexible conformational space.
    These features make vorinostat a representative test case for evaluating the capability of StoL.

    In the following, we present the initial geometries generated by StoL and the subsequent DFT optimization to illustrate the complete workflow.
    As summarized in Figure S4, the number of structures produced and retained at each stage is clearly delineated.

\subsubsection{StoL Generation Performance for the Illustrative Example}

    To demonstrate StoL's capability in handling a complex and flexible molecule, we applied the framework to vorinostat, following each step from fragment decomposition to full 3D structure reconstruction.

    The process begins with decomposing the target molecule into several fragments.
    In this case, 19 bonds were identified between heavy atoms, representing the molecular skeleton.
    Among them, eight bonds were excluded from cleavage, including two double bonds and six located within ring systems.
    Following this screening, the algorithm selected a set of three chemically valid fragments, as illustrated in Figure~\ref{fig:example}.(b).
    This fragment set satisfies all selection criteria: \textit{Frag1} and \textit{Frag2} share six heavy atoms (C2, C3, C4, C5, C6 and C7), while \textit{Frag2} and Frag3 share four (C7, C8, O9 and N10).
    Together, these overlapping regions ensure structural continuity across fragments, allowing \textit{Frag1}, \textit{Frag2}, and Frag3 to be seamlessly merged to reconstruct the complete molecular structure.
    This example demonstrates that the fragmentation algorithm effectively balances chemical validity and completeness while minimizing redundant overlaps between adjacent fragments.

    Next, the three fragment SMILES strings are passed into the trained StoL model to generate their 3D conformations.
    Taking \textit{Frag1} (\url{O=C(CCCCCC)NO}) as an example, 50 candidate structures were generated.
    Among these, 47 conformations passed the basic cheminformatics checks, indicating the high reliability of the fragment-level generative process.
    To further reduce redundancy and ensure diversity, the remaining structures were processed with dimensionality reduction followed by K-centroids clustering.
    In the given example (see Figure~\ref{fig:example}.(c)), 10 representative conformations were selected according to the geometrical distribution in the two-dimensional reduced space.
    These conformations reflect the intrinsic flexibility of \textit{Frag1}, highlighting the challenges of fully exploring the accessible chemical space of the entire molecule.
    The selected representative structures were then passed to the assembly stage, serving as the building blocks for reconstructing the full 3D molecular structure.

    Following fragment-level generation, representative conformations were assembled into complete vorinostat structures using the StoL assembly procedure.
    Several hundred geometries successfully passed the cheminformatics validation checks.
    From these, 100 representative 3D structures were selected using dimensionality reduction and clustering methods.
    For comparison, an additional 100 3D structures were generated using RDKit \cite{rdkit}, a widely adopted cheminformatics toolkit for molecular conformation generation.
    In the dimensionality-reduced space in Figure~\ref{fig:example}.(d), StoL-generated structures exhibit a much broader and more evenly distributed coverage, whereas those generated by RDKit cluster densely within a narrower region.
    This broader distribution indicates that StoL explores a wider and more diverse conformational space.
    The representative structures shown further demonstrate that the assembly protocol effectively maintains chemical validity while integrating flexible fragment conformations.

    To further evaluate the geometric accuracy of StoL, we directly compared the structures generated by StoL with those obtained from DFT optimization using RDKit-generated initial guesses.
    The 100 RDKit-generated conformations obtained above were optimized at the B3LYP/6-31G*/BJD3 level.
    Given the high complexity of the conformational space, only conformers differing by more than 1 kcal/mol in energy and by more than 1 \AA\ in scaffold RMSD were retained, resulting in 10 representative conformers and all of these structures should be accurate at the quantum chemistry level.
    We then compared these with the 100 representative StoL-generated structures.

    Here, we computed the scaffold RMSD between the ten representative RDKit+DFT conformers and the those 100 StoL-generated structures, resulting in a ($10 \times 100$) RMSD matrix.
    The minimal RMSD values in each row, representing the best reconstruction of each RDKit+DFT conformer, are summarized in Table~S5.
    As shown, StoL successfully reproduces all reference conformers with correspondingly small RMSD values, indicating that each target structure can be well reconstructed.
    However, because multiple RMSD values make it difficult to assess the overall quality of the generated structures, we introduced a Boltzmann-weighted RMSD (BRMSD) metric to provide a more comprehensive evaluation of conformational similarity across the chemical space.
    The assembled StoL conformations achieved an average BRMSD of 0.76~\AA\ relative to the reference set, indicating excellent structural consistency.
    Notably, unlike one-step generative models, the StoL framework never observes the complete molecular structure during training.
    Considering the substantial flexibility of vorinostat, this close agreement demonstrates that StoL accurately reconstructs 3D geometries and generates conformations comparable in quality to those produced by state-of-the-art one-step models. \cite{xuGeoDiffGeometricDiffusion2022,ganeaGeoMolTorsionalGeometric2021}

    Overall, this example highlights that StoL not only ensures chemical plausibility at the fragment level but also reconstructs accurate and diverse full-molecule geometries.
    Compared with RDKit, StoL offers substantially broader conformational coverage, demonstrating its effectiveness for systematic and exhaustive exploration of molecular conformational space.

\subsubsection{DFT Optimization of the Illustrative Exapmle}

\begin{figure}[H]
    \centering
    \includegraphics[width=\linewidth]{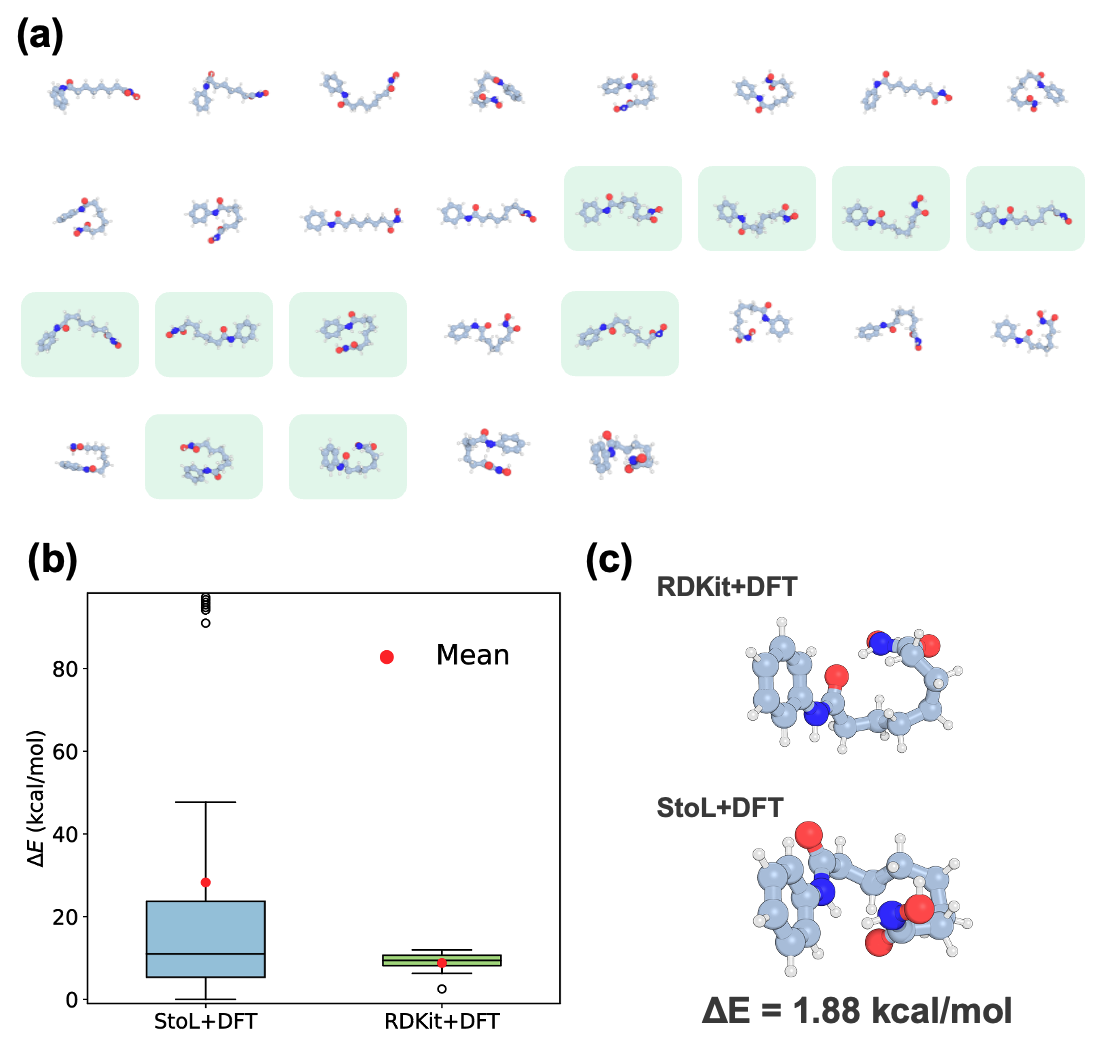}
    \caption{
        DFT evaluation of vorinostat conformations.
        (a) 29 representative conformations derived from DFT calculations using StoL-generated initial guesses, arranged in descending order of energy, with structures highlighted in light green indicating those obtained from both RDKit and StoL initial guesses.
        (b) Box plot of relative energies ($\Delta E$) referenced to the lowest-energy structure, comparing conformers generated from RDKit+DFT (StoL25-init dataset) and StoL+DFT (DFT calculation based on StoL-generated initial structures).
        (c) Comparison of the lowest-energy conformers obtained by the two methods, showing that the StoL+DFT conformer is more stable, with an energy 1.88 kcal/mol lower than the RDKit+DFT conformer.
        All molecular geometries were visualized via PyMOL. \cite{pymol}
    }
    \label{fig:example_dft}
\end{figure}

    To assess the quality of conformations generated by StoL, we conducted a comparative evaluation on vorinostat (\url{O=C(CCCCCCC(=O)Nc1ccccc1)NO}) after DFT optimization.
    The 100 geometries generated by StoL in the previous step were also optimized at the B3LYP/6-31G*/BJD3 level.
    After filtering to retain only conformers differing by more than 0.6 kcal/mol in energy and by more than 1 \AA\ in scaffold RMSD, 29 distinct conformations were obtained.
    All 10 conformers derived from RDKit were found to be included within this StoL-derived set.
    Figure~\ref{fig:example_dft}.(a) presents the 29 representative StoL+DFT conformers arranged in descending order of energy, with structures highlighted in light green indicating those shared with the RDKit+DFT results.

    The comparison reveals a clear difference in conformational coverage.
    RDKit-generated initial guesses lead to DFT-optimized structures clustered within a narrow energy region, indicating limited exploration of the conformational landscape.
    In contrast, StoL-generated initial structures yield a broader and more diverse set of DFT minima, spanning both high- and low-energy states.
    As shown in Figure~\ref{fig:example_dft}.(b), the energy distribution of StoL+DFT conformers extends far beyond that of RDKit+DFT, demonstrating StoL's superior ability to sample a wider range of local minima.

    Notably, the lowest-energy StoL+DFT conformer is 1.88 kcal/mol more stable than the most stable RDKit+DFT conformer (Figure~\ref{fig:example_dft}.(c)), underscoring StoL's capability to access more favorable and physically relevant structures.
    This broader energy spectrum and improved stability arise from StoL's fragment-based, LEGO-inspired design, which systematically reconstructs molecules from diverse fragment combinations rather than relying on a single-step, rule-based embedding.
    Collectively, these results highlight the robustness of StoL framework in generating chemically valid, energetically optimized, and structurally diverse conformations that more comprehensively capture the molecule's accessible conformational space.

\subsection{Overall Performance Analysis of StoL}

    To comprehensively evaluate StoL on realistic and chemically diverse systems, we constructed the StoL25-init benchmark dataset using Gaussian 16 \cite{g16}.
    The dataset comprises 200 drug-like molecules containing 16-25 heavy atoms (C, N, O, and F) selected from the ChEMBL database, \cite{gaulton2012chembl}, with further details provided in the Supplementary Information.
    For each molecule, 100 initial 3D conformations were generated using RDKit.
    Ten randomly selected conformations were subsequently optimized at the DFT B3LYP/6-31G*/BJD3 level.
    As shown in Figure~\ref{fig:stol_gen}(a), this procedure yielded a total of 1,691 unique geometries.
    Importantly, StoL25-init was constructed solely from the ChEMBL database because it is used for evaluation rather than model training.
    Notably, these molecules are significantly larger and more flexible than those in the training set, making direct diffusion-based conformer generation highly challenging.
    This benchmark thus provides a stringent test of StoL's scalability and transferability, demonstrating how fragment-level learning enables accurate reconstruction of large, flexible, and chemically complex molecules.
    Further details of dataset construction are provided in the Supplementary Information.

    In this section, we further evaluate StoL-generated structures from two complementary perspectives across the entire StoL25-init dataset: the quality of the initial geometries produced by StoL and the corresponding DFT-optimized conformations.

\subsubsection{Generation Performance of the StoL Framework}
    We first assess StoL's performance from multiple complementary perspectives, including fragmentation efficiency, training behavior, geometric fidelity, and molecular assembly accuracy.

\begin{figure}[H]
    \centering
    \includegraphics[width=0.8\linewidth]{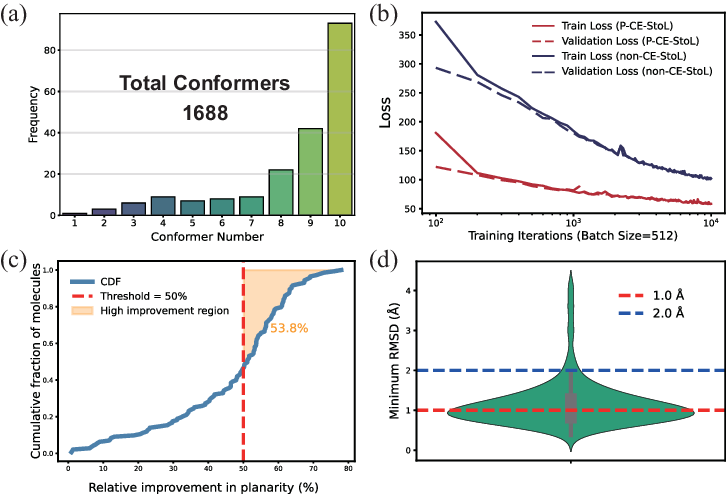}
    \caption{
    (a) Conformer distribution in the StoL25-init dataset. Each structures was obtained by DFT optimization of 10 RDKit-generated initial guesses.
    (b) Training loss curves comparing the P-CE-StoL and non-CE-StoL model. Red lines denote Train Loss (P-CE-StoL) and Validation Loss (P-CE-StoL), while blue lines denote Train Loss (non-CE-StoL) and Validation Loss (non-CE-StoL), plotted across training iterations.
    (c) Cumulative distribution of relative planarity improvement, with the x-axis showing the percentage increase of P-CE-StoL over CE-StoL and the y-axis showing the fraction of molecules with improvement less than or equal to that percentage. The yellow-shaded region highlights molecules with improvement greater than 50\%.
    (d) Violin plots of BRMSD values for generated structures with respect to reference conformations in StoL25-init dataset, with the red dashed line marking 1 \AA\ and the blue dashed line marking 2 \AA.
    }
    \label{fig:stol_gen}
\end{figure}

\subsubsubsection{\textbullet~Fragmentation}
\newline
    The StoL framework begins with the fragmentation of molecular SMILES strings, where an automated fragmentation protocol enables rapid and systematic decomposition.
    Valid fragments satisfying the defined criteria were successfully generated for all 200 benchmark molecules within seconds, demonstrating the efficiency and effectiveness of the fragmentation algorithm.
    Representative examples of the fragmentation process are provided in the Supplementary Information.
    In addition, nearly all the obtained fragments are novel and not included in the training set, confirming that the selected benchmark dataset provides an appropriate and unbiased basis for evaluating the model's generalization capability.

\subsubsubsection{\textbullet~Efficiency of chemistry-enhanced training}
\newline
    The training process plays a decisive role in determining the fidelity of StoL-generated molecular geometries.
    To assess the impact of the CE training strategy, the performance of P-CE-StoL was compared with that of non-CE-StoL under identical training conditions.
    As shown in Figure~\ref{fig:stol_gen}.(b), P-CE-StoL exhibited a markedly faster decline in training loss and achieved lower final loss values within the same number of epochs.
    This improvement confirms that embedding chemical knowledge into the training process substantially enhances data efficiency, allowing the model to extract richer structural information from limited data and converge more rapidly.
    Such efficiency gains are particularly valuable for scaling molecular generative models to larger and more complex chemical systems.

\subsubsubsection{\textbullet~Planarity preservation in aromatic systems}
\newline
    Planarity serves as a stringent indicator of geometric realism, especially for aromatic compounds where deviations often lead to physically implausible structures.
    Figure~\ref{fig:stol_gen}.(c) highlights that P-CE-StoL significantly improves planarity relative to CE-StoL, with
    more than half of aromatic molecules showing over a 50\% enhancement.
    This result underscores the necessity of incorporating a planarity-aware loss, which effectively constrains the model to maintain correct aromatic ring geometries.
    Such explicit geometric regularization is crucial for ensuring the chemical plausibility of generated molecular structures.

\subsubsubsection{\textbullet~Molecular assembly and structural fidelity}
\newline
    StoL reconstructs complete molecular geometries through fragment assembly and cheminformatics validation following the denoising process.
    Using the P-CE-StoL model, all 200 molecules in the StoL25-init dataset were successfully assembled and passed chemical validity checks.
    In contrast, using fragment structures generated by the non-CE-StoL model resulted in successful reconstruction for only 79 molecules.
    This striking contrast underscores the effectiveness of the CE training and planarity-aware constraints in preserving chemical integrity and geometric coherence during molecular reconstruction.

    As illustrated in Figure S5, StoL-generated conformations display a broader distribution of geometric space compared with those from RDKit.
    This observation reflects StoL's ability to capture the intrinsic flexibility and conformational diversity of complex molecules rather than converging to overly constrained local minima.

    A quantitative comparison with DFT-optimized structures further confirms StoL's high structural fidelity.
    As shown in Figure \ref{fig:stol_gen}.(d), most assembled conformations exhibit low Boltzmann-weighted RMSD (BRMSD) values, with over half (101 molecules) below 1.0 \AA \ and only 9 highly flexible molecules exceeding 2.0 \AA\ (listed in Table~S9).
    These results demonstrate that StoL consistently reproduces accurate 3D geometries across a wide range of molecular flexibilities.
    The few less accurate cases are primarily associated with oxygen-rich fragments (more than four O atoms), which constitute only 2.1\% of the training database.
    This deviation likely stems from data imbalance rather than model deficiency, suggesting that expanding fragment diversity in future datasets would further enhance StoL's generalization to underrepresented chemical motifs.


\subsubsection{DFT Optimization accross Whole Dataset}

\begin{figure}[H]
    \centering
    \includegraphics[width=0.8\linewidth]{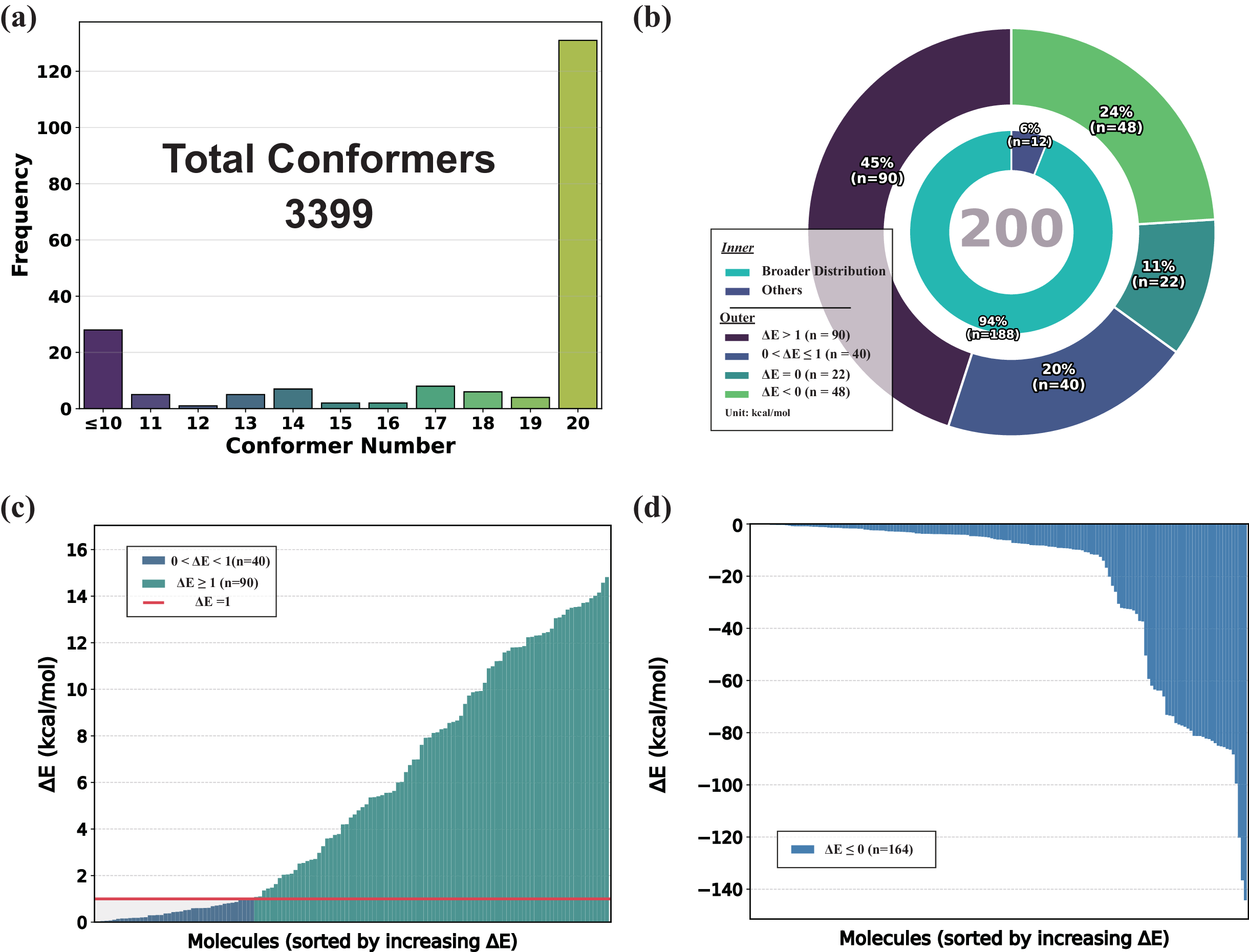}
    \caption{
    (a) Distribution of conformers after DFT optimization using 20 representative StoL-generated initial structures.
    (b) Donut chart showing the energy differences ($\Delta E^{min} = E^{min}_{\text{RDKit+DFT}} - E^{min}_{\text{StoL+DFT}}$) between the lowest-energy conformers obtained from the RDKit- and StoL-initialized DFT workflows for 200 molecules (in kcal/mol).
    (c) Bar chart of $\Delta E^{min}$ values for molecules where the StoL+DFT workflow yields a lower-energy conformer, sorted by ascending energy difference (in kcal/mol).
    (d) Bar chart of the energy difference between the conformation with highest energy, where $\Delta E^{max} = E^{max}_{\text{RDKit+DFT}} - E^{max}_{\text{StoL+DFT}}$.
    Only molecules that StoL+DFT workflow yields a higher-energy conformer are shown, sorted by ascending energy difference (in kcal/mol).
    }
    \label{fig:count}
\end{figure}

    We further assessed the chemical validity and practical utility of conformations generated by StoL across all molecules in the StoL25-init dataset.
    Specifically, DFT optimizations were performed at the B3LYP/6-31G*/BJD3 level using 3D geometries generated for these 200 drug-like molecules.
    Initially, 20 representative conformers were selected for each of the 200 drug-like molecules through dimensionality reduction and clustering, yielding a total of 4,000 structures.
    To ensure a fair comparison with structures StoL25-init dataset obatained from RDKit-generated intial guess, DFT optimizations at the B3LYP/6-31G*/BJD3 level were first performed on half of these structures (10 conformers per molecule, 2,000 in total).
    Detailed comparisons are provided in the Supporting Information.
    The remaining 2,000 conformers were subsequently optimized under the same conditions, showing consistent trends with the first half subset.
    Therefore, the main text discusses the results of all 4,000 DFT-optimized structures collectively, confirming the high reliability of StoL as an efficient and chemically valid initial conformer generator.
    All optimized structures were also verified to ensure that the molecular identities remained unchanged.

    Ultimately, as shown in Figure~\ref{fig:count}.(a), 3,399 conformations were successfully obtained, reflecting the high reliability of StoL as an efficient initial conformer generator.
    Among the optimized structures, 2,275 unique conformers not present in the original dataset were identified.

    In our energetic analysis, as shown in Figure~\ref{fig:count}.(b), StoL successfully identifies lower-energy conformers for 130 molecules (65\%), with 90 molecules (45\%) exhibiting energy improvements greater than 1 kcal/mol relative to the corresponding reference structures in the StoL25-init dataset.
    This indicates that StoL not only reconstructs known conformations but also discovers more energetically favorable ones that were previously inaccessible.StoL-generated structures further exhibit a broader energy landscape, with 188 out of 200 molecules showing wide energy spans.
    Further analysis of both low- and high-energy conformations reveals consistent findings.
    As illustrated in Figure~\ref{fig:count}.(c), the $\Delta E$ distribution of these 130 molecules confirms StoL's consistent ability to locate lower-energy states across diverse chemical systems.
    Moreover, Figure~\ref{fig:count}.(d) reveals that 164 molecules also possess higher-energy conformations, with many extending tens of kcal/mol above the minimum, reflecting StoL's comprehensive exploration of both low- and high-energy regions of the potential energy surface.
    Overall, these results collectively highlight StoL's ability to extend beyond traditional generative approaches, not only reproducing known low-energy structures but also identifying new, physically meaningful conformations that expand the accessible chemical space.
    Additional examples are provided in the Supplementary Information.

    These findings highlight the ability of StoL not only to reproduce known conformers but also to broaden the accessible energy landscape, enhancing sampling diversity and enabling the discovery of energetically more favorable states absent from the reference dataset.

    As an additional contribution of this work, we incorporated all newly identified conformers into the StoL25-init dataset and expanded it to form the StoL25 dataset. The updated dataset now includes 200 molecules with 3,948 conformers and is publicly available (https://langroup.site/StoL25/) to support further research.
    This expanded dataset provides a richer and more diverse conformational space for these drug-like molecules and serves as a valuable benchmark for future development and evaluation of conformer generation methods.
    The dataset will be continuously maintained and updated as new conformers and molecular systems are identified.

\subsection{Limitations of Current Methods}
\label{sec:lim}

    Despite the promising performance of the StoL framework, several inherent limitations highlight important avenues for further improvement.

    A notable limitation of the current StoL framework arises from its training dataset, which is restricted to fragments with no more than 10 heavy atoms, thereby impeding the handling of molecules featuring multiple or fused ring systems.
    This constraint prevents the generation of valid fragments for large polycyclic structures without violating ring bonds, potentially compromising the fragmentation process.
    To mitigate this, expanding the training dataset to incorporate representative multi-ring motifs, coupled with ring-system-specific rules for training, fragmentation, and assembly, would enhance performance.
    Alternatively, integrating a dedicated sub-model to explicitly enforce ring closure constraints could further improve StoL's capability in addressing complex ring systems.

    Another limitation arises from the model's fragment-by-fragment generation strategy, which primarily captures local structural correlations.
    For large flexible molecules where long-range noncovalent interactions dominate conformational preferences, this localized approach may fail to accurately reproduce global geometry.
    A feasible solution is to integrate an auxiliary correction network that iteratively refines assembled structures by accounting for long-range electrostatic and dispersion interactions.
    Embedding distance-dependent interaction potentials or graph-based message passing between nonadjacent fragments could further improve structural coherence.

    Despite its advantages in simplicity and computational efficiency, the current StoL framework inevitably sacrifices some robustness due to its single-model design.
    While ensemble diffusion strategies could enhance stability and diversity,\cite{kimDiffusionbasedGenerativeAI2024} they also entail a considerable sampling cost.
    In this context, introducing faster or more efficient sampling strategies offers a promising direction to further accelerate inference and broaden the applicability of StoL to large molecular systems.
    Such developments would strengthen StoL's balance between accuracy and efficiency, making it more suitable for large-scale molecular discovery tasks.

\section{Conclusions}

In this work, we introduced StoL, a chemistry-enhanced diffusion-based framework for efficient 3D conformation generation of large molecules from only SMILES inputs.
StoL adopts a modular, LEGO-inspired design consisting of three sequential stages: first, molecular structures are decomposed into chemically valid fragments; second, each fragment is locally reconstructed via diffusion-based generation; and third, the fragments are assembled into complete molecules.
This strategy enables StoL to explore the conformational landscape of large, flexible systems in a scalable, data-efficient, and chemically faithful manner.

Compared to conventional generative paradigms that require large molecule-specific datasets, StoL fundamentally shifts the data requirement toward small molecular fragments.
This fragment-centric design allows both training and inference to be performed within a compact and chemically meaningful space, significantly enhancing transferability and scalability.
By embedding chemical principles across fragmentation, training, filtering, and assembly, StoL maintains high geometric fidelity and ensures that generated structures are not only geometrically plausible but also chemically valid.
In this sense, StoL bridges the gap between deep generative modeling and chemical realism.

Comprehensive evaluation on the StoL25-init dataset underscores the robustness of the framework.
First, StoL-generated conformers exhibit significantly broader coverage of conformational space than those produced by rule-based methods such as RDKit.
Second, when compared with RDKit+DFT results, StoL-based conformers show excellent performance on capture whole space, as quantified by the chemistry-informed BRMSD metric, confirming their chemical fidelity.
Third, DFT refinement of StoL-generated conformers (StoL+DFT) further expands the accessible energy landscape and, in several cases, identifies lower-energy minima than RDKit+DFT, demonstrating StoL's ability to access deeper potential energy basins.
Together, these findings show that StoL not only enhances conformational diversity but also improves the exploration of the energetic landscape.

Overall, StoL establishes a new paradigm for integrating chemical knowledge into generative modeling.
By constructing large-molecule conformations from fragment-level minima, it enables systematic and interpretable exploration of potential energy surfaces, providing a robust foundation for studies of molecular dynamics, reactivity, and property prediction.
Looking ahead, StoL can be further strengthened by incorporating additional physical constraints and expanding high-quality datasets to improve accuracy, generalization, and applicability.
Ultimately, StoL exemplifies how infusing chemical insight into diffusion-based generation can fundamentally reshape the way molecular structures are learned, assembled, and understood.


\section{Methods}
\subsection{Diffusion Model in StoL}

Within the StoL framework, we employ a diffusion model to generate 3D conformations of molecular fragments, conditioned on the molecular graph $ G $ derived from SMILES strings.
Diffusion models \cite{songScoreBasedGenerativeModeling2021, hoDenoisingDiffusionProbabilistic2020} operate through a two-phase process: a forward (noise-addition) phase and a reverse (denoising) phase.
In the forward phase, Gaussian noise is incrementally added to the atomic coordinates of ground-truth fragment structures over a series of discrete timesteps, gradually transforming the data into pure noise.
The reverse phase trains a neural network to invert this process, learning to reconstruct the original conformation by progressively denoising the perturbed input.
Grounded in Markov chain formulations, this approach enables stable and high-quality sampling, offering a robust alternative to GANs and VAEs.
For a detailed theoretical derivation, refer to the Supplementary Information and foundational works \cite{hoDenoisingDiffusionProbabilistic2020, songScoreBasedGenerativeModeling2021, xuGeoDiffGeometricDiffusion2022, kimDiffusionbasedGenerativeAI2024}.

In our implementation, the forward process corrupts atomic positions in Cartesian space according to a predefined noise schedule.
In the forward process, noise is incrementally added to the atomic positions of 3D fragment structures from a database over discrete timesteps.
The reverse process then learns to denoise this noisy state, guided by a neural network, to recover the original conformations.
Training minimizes a variational lower bound on the negative log-likelihood, with the loss function adapted from GeoDiff \cite{xuGeoDiffGeometricDiffusion2022} to accommodate fragment-level generation and chemical conditioning via $ G $.


\subsection{Chemistry-Enhanced Training}

Training directly in Cartesian space makes the denoising model sensitive to arbitrary molecular orientations and atom orderings, which can obscure structural learning. To improve data efficiency, a CE training strategy was implemented, consisting of two sequential phases: a data-driven phase and a chemistry-enhanced phase, as illustrated in Figure~\ref{fig:fig2}.(b).

In the data-driven phase, the model reconstructs molecular scaffolds from fully disordered states in a standard diffusion manner.
The chemistry-enhanced phase introduces chemical priors to refine structural alignment and improve correspondence between predicted and target geometries.
This phase consists of two substages: soft matching and hard matching.

\begin{itemize}
    \item  Soft matching: For each atomic element, a differentiable soft assignment between predicted and reference coordinates is computed using the Sinkhorn algorithm \cite{sinkhorn1967concerning}.
    \item Hard matching: As training stabilizes, the soft assignments are converted into fixed one-to-one mappings using the Gumbel-softmax method \cite{jangCategoricalReparameterizationGumbelSoftmax2017}, ensuring deterministic atom indexing for subsequent symmetry handling.
\end{itemize}

To remove non-essential spatial variance, all structures are centered at the origin, and rotational alignment is performed using the Kabsch algorithm \cite{kabsch1976solution, kabsch1978discussion}.
Mirror symmetry is handled by evaluating both original and z-axis-reflected configurations, retaining the alignment with the lower RMSD.
A planarity penalty term is added to preserve the geometry of aromatic rings and multiple bonds, defined by deviations from fitted ring planes.

Hydrogen atoms are explicitly included, with element-wise loss weights adjusted by phase:
\begin{itemize}
    \item  Data-driven phase: fixed at 0.5 to prioritize heavy-atom scaffold learning.
    \item  Soft matching: linearly increased from 0.5 to 1.0.
    \item  Hard matching: fixed at 1.0 to ensure full chemical consistency.
\end{itemize}

This staged procedure separates global geometric reconstruction from local chemical refinement, enhancing training stability and reducing data requirements.
Detailed parameter settings and implementation notes are provided in the Supplementary Information.

\subsection{Fragmentation}
The input SMILES string of a given molecule is first partitioned into several fragments based on the following rules.
(1) Each fragment contains a moderate number of heavy atoms (typically 6-10, depending on the training set).
(2) Adjacent fragments overlap by at least three heavy atoms.
(3) Fragmentation is restricted to breaking only single bonds outside of rings.
(4) The set of fragments must collectively cover the entire molecule.
In the StoL framework, fragmentation can be performed automatically using in-house scripts or manually guided, as detailed in the Supplementary Information.

\subsection{Fragment Filter}
For each fragment SMILES, candidate conformations are generated by the trained diffusion model through reverse diffusion.
Before assembling these fragments, it is important to verify whether the fragments are generated correctly to reduce computational resources during the assembly process.
First, the generated conformations are transformed into a molecular graph and compared with the SMILES-generated structure to validate the molecular connectivity and bond lengths.
Second, the flatness of the aromatic rings and the structure of double/triple bonds are also verified.
Conformations meeting these criteria are further processed using unsupervised ML techniques, that is, UMAP algorithm \cite{sainburg2021parametric} and K-Medoids clustering method \cite{PARK20093336}, see Supplementary Information.
The cluster centers resulting from this analysis are then chosen for the final assembly stage.

\subsection{Fragment Assembly}

Fragments are sequentially paired and assembled through exhaustive enumeration of candidate combinations.
The critical step in assembling conformations from two fragments lies in identifying their common structural region.
Given that chemical symmetries are incorporated into the model training, we utilize an index-tracking trick to address this challenge.
This method involves tracking the indices of each atom from the original molecule throughout all operations, facilitated by graph comparison.
Consequently, the common part can be easily identified.
Subsequently, heavy atoms in the shared region are aligned using the Kabsch algorithm (with mirroring considered), and fragment pairs with RMSD below a small threshold (e.g., 0.3 Å) are retained.
Following this, the common region of the latter fragment is removed.
During bond rejoining, redundant hydrogens are removed by comparing the cosine similarity between the target bond vector and adjacent \ch{H-X} vectors.
After iteratively processing all fragment pairs in sequence, multiple conformations of the original large molecule can be generated.

\subsection{Cheminformatics Check}

\begin{table}[htbp]
    \centering
    \caption{Cheminformatics validation checks}
    \begin{tabular}{p{0.25\textwidth} p{0.55\textwidth}}
    \toprule
    \textbf{Validation Check} & \textbf{Description} \\
    \midrule
    File Loading / SMILES Transformation & Ensures proper loading of conformations from files and ensures their accurate conversions to SMILES strings. \\
    Sanitization & Verifies the chemical validity of the molecular structure by correcting or flagging issues such as invalid valences or aromaticity. \\
    Formula Check & Confirms that the molecular formula derived from the generated conformation matches the expected composition. \\
    Bond Connectivity & Assesses the correctness of bond connections, ensuring that the topology aligns with target molecules. \\
    Bond Lengths & Evaluates the geometric accuracy of bond distances, comparing them against standard values for the respective atom pairs. \\
    Bond Angles & Analyzes the angular geometry between bonded atoms to ensure compliance with expected chemical angles. \\
    Internal Steric Clash & Ensure that no steric interactions occur during the assembly of fragments.\\
    \bottomrule
    \end{tabular}
    \label{tab:validation_checks}
\end{table}

    Drawing inspiration from PoseBusters, \cite{buttenschoen2024posebusters} a widely recognized tool for assessing protein-ligand docking, we integrated several cheminformatics validation checks into the StoL codebase to evaluate the quality of generated conformations for large molecules, as shown in Table~\ref{tab:validation_checks}.
    These comprehensive validations enhance the reliability of the generated conformations by addressing both chemical accuracy and spatial plausibility, bridging the gap between computational generation and valid molecular structures.

\section{Data Availability}

The QCDGE dataset used as training data in this work is available at 
\url{https://langroup.site/QCDGE/} (ref.~\citenum{zhuQuantumChemistryDataset2024}). 
The StoL25 dataset is accessible at \url{https://langroup.site/StoL25/}.

\section{Code Availability}
Our source codes are publicly available at the Github repository (\url{https://github.com/Yifei-Zhu/StoL.git}).

\section*{Supporting Information Available}
    Several relevant information:
    Computational details of the diffusion model, chemistry-enhanced training, equivariance analysis, the Sinkhorn and Gumbel-softmax algorithms, as well as other related methods; and additional results and discussions including illustrative example (vorinostat), StoL generation, DFT optimization, and information of molecules in StoL25 dataset.

\section*{Author Information}
\subsection*{Corresponding Authors}
    E-mail: zhenggang.lan@m.scnu.edu.cn

\subsection*{Notes}
    The authors declare no competing financial interest.

\begin{acknowledgement}
    The authors express sincerely thanks to the National Natural Science Foundation of China (No. 22333003 and 22361132528) for financial support.
    Some calculations in this paper were done on SunRising-1 computing environment in Supercomputing Center, Computer Network Information Center, CAS.
\end{acknowledgement}

\begin{tocentry}
\includegraphics[width=8.4cm, keepaspectratio]{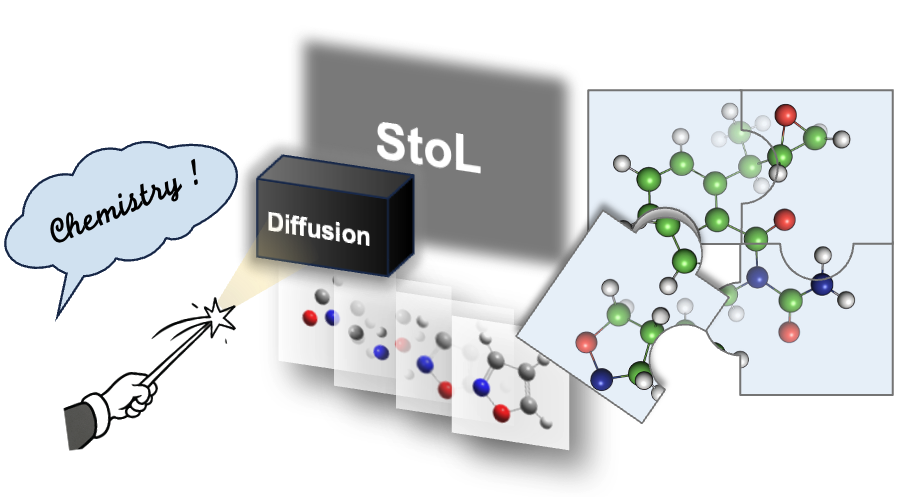}
\end{tocentry}

\bibliography{stol}
\end{document}